\documentclass{article}
\usepackage{spconf,amsmath,graphicx}
\usepackage{amsmath,graphicx}
\usepackage{xspace}
\usepackage{siunitx}
\usepackage{bm}
\usepackage{hyperref}
\usepackage[justification=centering]{subcaption}
\usepackage{fancyhdr}
\usepackage{booktabs}
\usepackage{multirow}
\usepackage{graphicx}

\usepackage{enumitem}
\setlist{nosep, leftmargin=14pt}

\usepackage{mwe} %
\usepackage{amsfonts}
\usepackage{nicefrac}
\usepackage{amsmath}
\usepackage{bm}

\graphicspath{{images/}}

\usepackage{xcolor}

\newcommand{\tref}[1]{Table~\ref{#1}}
\newcommand{\norm}[1]{\left\lVert#1\right\rVert}

\fancypagestyle{specialfooter}{%
  \fancyhf{}
  
  \fancyfoot[C]
  {\textbf{$\copyright$ 2021 IEEE. Personal use of this material is permitted. Permission from IEEE must be obtained for all other uses, in any current or future media, including reprinting/republishing this material for advertising or promotional purposes, creating new collective works, for resale or redistribution to servers or lists, or reuse of any copyrighted component of this work in other works.}
  }
}
\title{GORDA: Graph-based ORientation Distribution Analysis of SLI scatterometry Patterns of Nerve Fibres}

\name{Esteban Vaca$^{1,2}$ \qquad Miriam Menzel$^{1}$  \qquad Katrin Amunts$^{1,3}$ \qquad Markus Axer$^{1}$   \qquad  Timo Dickscheid$^{1,2}$}%
\address{
	$^{1}$ Institute of Neuroscience and Medicine (INM-1), Research Centre Jülich, Germany\\
	$^{2}$ Helmholtz AI, Research Centre Jülich, Germany\\
	$^{3}$ C\'{e}cile \& Oskar Vogt Institute for Brain Research, University Hospital Düsseldorf, Germany\\
	$^{4}$ Institute of Computer Science, Heinrich-Heine-University Düsseldorf, Germany
}

\begin{document}
\maketitle
\begin{abstract}
	Scattered Light Imaging (SLI) is a novel approach for microscopically revealing the fibre architecture of unstained brain sections. The measurements are obtained by illuminating brain sections from different angles and measuring the transmitted (scattered) light under normal incidence.
	The evaluation of scattering profiles commonly relies on a peak picking technique and feature extraction from the peaks, which allows quantitative determination of parallel and crossing in-plane nerve fibre directions for each image pixel. However, the estimation of the 3D orientation of the fibres cannot be assessed with the traditional methodology. 
    We propose an unsupervised learning approach using spherical convolutions for estimating the 3D orientation of neural fibres, resulting in a more detailed interpretation of the fibre orientation distributions in the brain.  
\end{abstract}
\begin{keywords}
	Human Brain, Fibre Architecture, Unsupervised Learning, CNNs, Spherical Convolution.
\end{keywords}

\vspace*{-.5\baselineskip}
\section{Introduction}
\label{sec:intro}

The nerve fibre architecture of the brain is a complex system comprised of billions of linked neurons~\cite{lent2012many}. Understanding the architecture of the nerve fibre network in the brain is essential for explaining functions of the brain and can be utilized for finding therapies for degenerative disorders ~\cite{aung2013diffusion}.

In recent years, several imaging methods have been developed to disentangle the brain's complex architecture, resolving the anatomical structure at multiple spatial resolutions.
Diffusion-weighted magnetic resonance imaging (DW-MRI) is the most prominent technique since it may be used on both in vivo and postmortem brains reaching millimetre scale in vivo subjects and resolutions up to 100 \textmu m in postmortem samples~\cite{beaujoin2018post}. However, DW-MRI resolution is not able to capture the structure of fibres with diameter ranging from $0.16$ to  9 \textmu m and an average of 1 \textmu m~\cite{liewald2014distribution}. Therefore, imaging techniques are required to assess the fibre architecture at microscopic resolutions in postmortem brains. Among those,  3D Polarized Light Imaging (3D-PLI)~\cite{axer2011high}  measures the three-dimensional orientation of nerve fibres in histological brain sections with in-plane resolutions of up to $1.3$ \textmu m, effectively reconstructing the joint orientation of fibres in a volume fraction. However, the exact computation of 3D orientations is affected by partial volume effects.

To better capture the fibre architecture of complex structures, Scattered Light Imaging ~\cite{menzel2021scattered, menzel2021scatterometry, reuter2020slix} has been successfully used to reveal fibre crossings in brain tissues. Nevertheless, to the best of our knowledge, light scattering pattern analysis has not yet been exploited to obtain a complete description of the 3D fibre orientation distribution function (fODF). For instance, ~\cite{menzel2021scatterometry} analyses the patterns applying a peak picking algorithm to the line profile generated by the integral along the radial axis of the scattering pattern, neglecting the complete structure of the scattering pattern and retrieving only the in-plane orientation of the fibres.

Here, we aim to capture the complete 3D-fODF from the scattering patterns. Since no groundtruth fODF is available for this type of data, we draw inspiration from ~\cite{elaldi2021equivariant} to perform unsupervised learning to predict the fODF in each pixel of the image. Since the scattering patterns cannot be reconstructed from linear operations such as convolution in diffusion MRI\cite{tournier2004direct}, we employ a response mixture model, which allows us to compare the input signal with a reconstructed model based on the estimated model fODF. We compare the performance of the proposed method with the results obtained by~\cite{Amunts2015}. Our main contributions are: 
\begin{enumerate}
    \item We present a graph-based orientation distribution analysis  (GORDA), an unsupervised spherical convolutional network, to reconstruct the fODF from light scattering patterns without the need for groundtruth labels. 

    \item A representation of scattering patterns on a spherical surface, enabling spherical convolutions, allowing to capture the spherical nature of fODF.
        
    \item A geometric  model of the response of  fibres at different inclinations and rotation angles, using the superposition characteristic to perform pattern mixture and reconstruct the scattering patterns. 
    
\end{enumerate}

\begin{figure*}[t]
	\includegraphics[width=\textwidth]{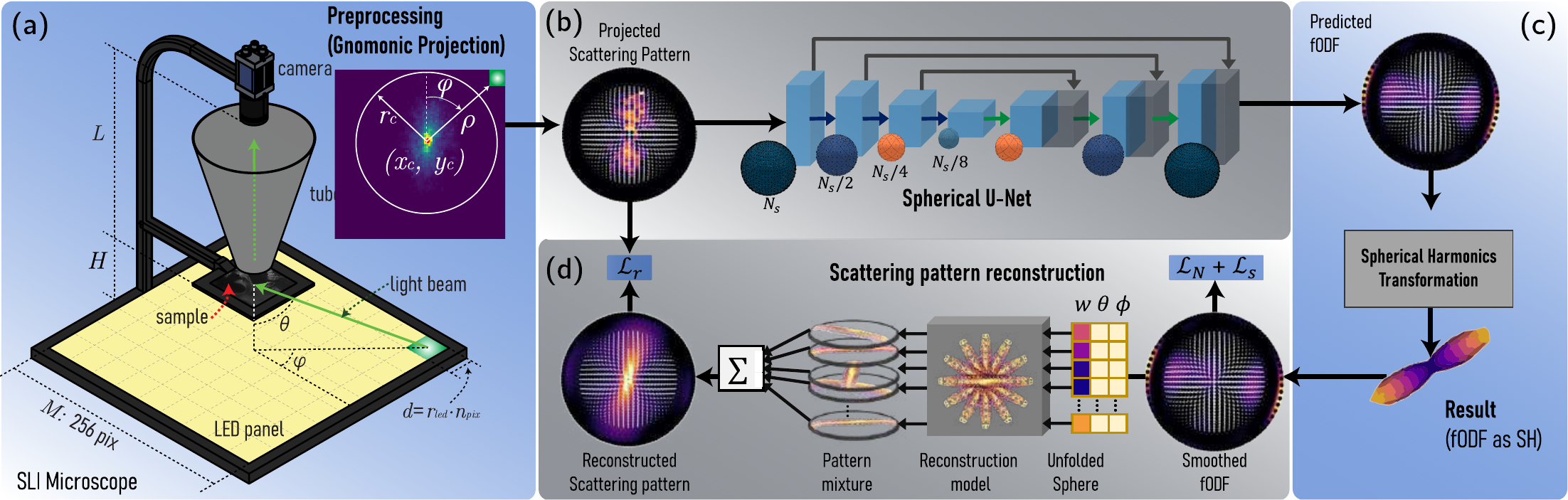}
	\caption{
		The proposed framework. (a) Measurement setup. The coordinates from the microscope with respect to the scattering pattern ($\phi$, $\theta$) are used to interpolate the scattering pattern to the HEALPix \emph{HEALpix} sphere~\cite{gorski2005healpix}.
        The parameters of the microscope are $H=13$ cm and $L=40$ cm, and $r_{\text{led}}= 1.8$ mm (b) A Spherical U-Net is used to reconstruct the fODF from the scattering pattern. The grid resolution of each sphere is given by the $N_s = N_{side}$ parameter of \emph{HEALpix}, (c) Real spherical harmonics transformation compresses the resulting fODF. (d) Inverse spherical harmonic transformation translates the compression effectively into smoothing the resulting fODF. The resulting sphere is unfolded and processed by a geometric reconstruction model. The response of all the orientations in the fODF is then used inside a weighted mixture model to reconstruct the complete scattering pattern. Finally, the obtained reconstruction is compared with the projected scattering pattern to calculate $\mathcal{L}_r$, and the smoothed fODF is used for the calculation of the sparsity $\mathcal{L}_r$ and non-negativity $\mathcal{L}_n$  loss. 
	}
	\label{fig:scheme}
\end{figure*}

\section{Methods}%
\vspace*{-.5\baselineskip}
\label{sec:methods}

An overview of our framework is depicted in Fig.\ref{fig:scheme}. The concept details and the design of each component of our framework are described in the following points.

\textbf{Unsupervised Learning.}
The development of the current framework takes advantage of the Finite-difference time-domain simulations of the brain’s nerve fibre architecture developed by~\cite{menzel2018finite}. Since we do not have access to any groundtruth or expert annotations for the measured scattering patterns, simulations constitute a valuable tool for describing their structure for different fibre populations. Nevertheless, one pitfall of the simulations presented in~\cite{menzel2018finite} remains in their complexity and expensive computation, which requires approximately 8K cpu-hours for the simulation of a single nerve fibre architecture on an HPC cluster.  This makes it \emph{theoretically} impossible to run the simulation simultaneously to training and increase the complexity to create a comprehensive database for all the combinations of nerve fibres.  

To cope with this limitation, we develop a geometric projection of an ellipsoid that formulates an approximate model of the response of the scattering patterns at different orientations and inclinations for a single nerve fibre.
Since~\cite{menzel2018finite, menzel2021scattered}  show the existence of a superposition property of the patterns, we perform a mixture of all possible projections in a sphere to generate a reconstruction of the input pattern. 
The reconstructed pattern can be compared with the input measurement to optimize our spherical CNN. Details of each component of our framework are described as follows:

\textbf{Preprocessing.}
The centre of the scattering patterns varies between image pixels because each image pixel represents a different position in the sample and thus a different position with respect to the centre of the LED (light-emitting diode) display. Determining the centre of the scattering pattern is crucial to convert plain scattering patterns to spherical coordinates. To find the centre, we refine the method reported in~\cite{menzel2021scatterometry}, which considers the pixel with maximum intensity as the scattering pattern’s centroid. We modify their approach by applying a Gaussian smoothing filter over the  scattering pattern before finding maximum intensity pixel. The filter reduces the effect of noise in the scattering patterns. 

Once the centre of the measurement $(x_c,y_c)$ is found, we perform a projection from the measurement space to a spherical system. Since the fODF is a function in the sphere that can be well modelled either as spherical harmonics (SH) or as spherical functions, we take advantage of the spherical sampling method \emph{HEALPix}~\cite{gorski2005healpix}. \emph{HEALPix} provides a subdivision of a spherical surface in which each pixel covers the same surface area as every other pixel.

To convert the scattering pattern, acquired with the measurement setup presented in Fig.  \ref{fig:scheme}.a, to the spherical space, an inverse gnomonic projection is applied. The projection requires the pixel distance to the centre point defined as $\Delta x = x-x_c$, and $\Delta y = y-y_c$, where $x,y$ are the pixel coordinates in the scattering pattern. The mapping from the measurement to the sphere is calculated as $\phi = \arctan(\frac{\Delta x}{\Delta y}) $  and  $\theta = \arctan(\frac{d}{H} \sqrt{ {\Delta x}^2 + {\Delta y}^2 }) $. We interpolate the values from the measurement sphere to the \emph{HEALPix} sphere using the angular distance between the surface of the spheres as described in~\cite{karimi2021learning}.

\textbf{Spherical Convolutional Network. }
Our framework employs the graph-based spherical convolution developed in \emph{DeepSphere}~\cite{perraudin2019deepsphere} as the basis of the spherical U-Net architecture  ~\cite{Ronneberger2015}. The main advantage of spherical convolution over traditional convolution is that rotation invariance~\cite{esteves2018learning, perraudin2019deepsphere}, reducing the need for intensive data augmentation. 
Furthermore, the convolutions of \emph{DeepSphere} are computationally efficient since they have a linear computational complexity with respect to the number of pixels. They are adaptable to signal analyses that partially cover the spherical surface, which is the case in scatterometry patterns for which the maximum angle of incidence is $\theta=60^\circ$.

The spherical convolutional network is based on the U-Net architecture~\cite{ronneberger2015u}, with operations defined on the  \emph{HEALPix} sphere, allowing for down-sampling and up-sampling the spherical grid resolution ($N_\text{pix}$) with a number of pixels $N_\text{pix}= 12 N_s^2,$  with $N_s \in \{ i \in \mathbb{N}^0 \; \vert \; i \bmod 2 = 0 \}$ which defines the number of divisions along the side of a base-resolution. The setup enables capturing the response of the signal in the whole sphere and reconstructing the fODF.

\begin{figure}[!tb]
\centering
	\includegraphics[width=0.4\textwidth]{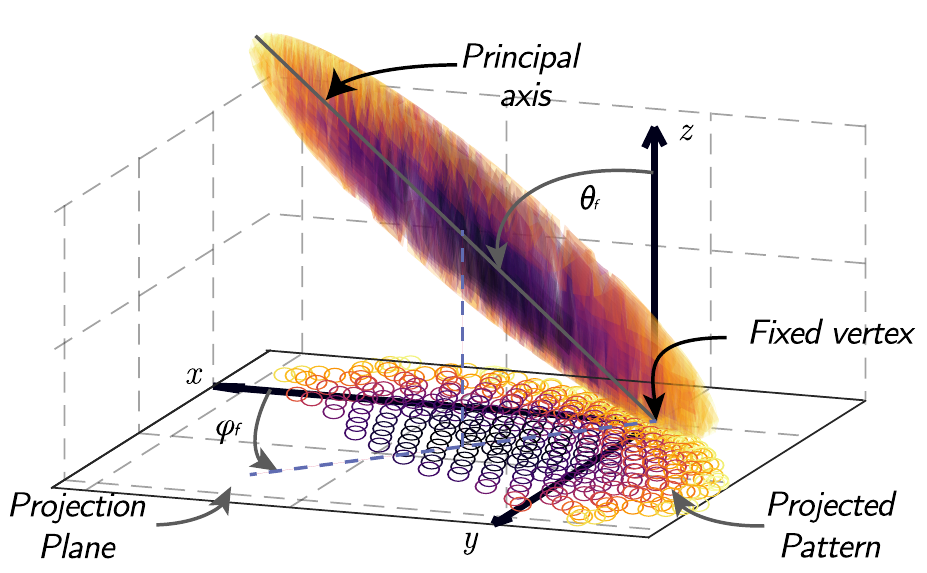}
	\caption{ Representation of the sphere used for reconstructing the scattering patterns, and its projection. The angles $\phi_f$ and $\theta_f$ modify the orientation of the sphere in the space.   
	}
	\label{fig:ellipsoid}
\end{figure}

\textbf{Scattering pattern reconstruction. }
To model inclination and rotation effects of the fibres in the scattering patterns, we take advantage of the simulations presented in~\cite{menzel2018finite}. The simulations show that the patterns rotate continuously with changes in the azimuthal angles of the fibres. Nevertheless, with fibre orientations switching out of plane due to increasing inclination angles, a ``C-shape" can be observed in the patterns. This effect is similar to the one observed when projecting an ellipsoid to a plane, and the ellipsoid is rotated in its azimuthal and zenith angles, keeping one of its co-vertices fixed. To represent these rotations, we define an ellipsoid as  
$(\bm{x}-\bm{x_c})^T\bm{R}^T\bm{\Lambda}\bm{R} (\bm{x}-\bm{x_c}) = 1$, where $\bm{x}$ is the position of each point in the ellipsoid, $\bm{x_c}$ is the centre of the ellipsoid, $\bm{\Lambda}$ is the diagonal matrix of its eigenvalues given by $[\alpha,1,1]$, with  $\alpha$ adjusting parameter the ``thickness" of the pattern, and   ${\bm{R}}$ the rotation matrix calculated from the azimuthal $\phi_f$ and zenith $\theta_f$ angles of the fibre as $\bm{R}= \bm{R}_y(\theta_f) \bm{R}_z(\phi_f)$. Fig. \ref{fig:ellipsoid} presents an overview of the proposed reconstruction model for a single fibre.

\textbf{Loss functions. } The proposed model is trained with respect to the characteristics of the scattering patterns and the estimated fODF. First, we define a reconstruction loss $\mathcal{L}_r$  to compare the whole input signal and the reconstructed scattering pattern. We follow the loss definition of~\cite{mhiri2021non} using a combination of $L_2$ loss and the Pearson correlation coefficient ($PCC$). Such combination captures the element-wise differences and  the overall correlation between  the input and reconstructed scattering pattern. The reconstruction loss is is defined as: 
\begin{equation}
	\mathcal{L}_r = \norm{ S- S_r }^2 + \lambda_r (1-PCC(S, S_r)) .
	\label{eq:reconstrucion_loss}
\end{equation}

Prior constraints are essential to regularize the training process of an unsupervised setting. In the case of the estimation of fODF,  sparsity priors are commonly used to recover the actual fibre orientation~\cite{auria2015structured} since the number of directions of interest in a pixel remains sparse. Our method is in line with the sparsity regularisation presented in~\cite{elaldi2021equivariant}, using a Cauchy distribution  to regulate the sparsity of the fODF as:

\begin{equation}
	\mathcal{L}_s = \lambda_s \sum^N_{i=1}{\log\left(1+\frac{\text{fODF}_i^2}{2\sigma^2_s}\right)} ,
	\label{eq:sparsity_loss}
\end{equation}

\noindent where $\lambda_s$ is the regularization coefficient and  $\sigma^2_s$ controls the sparsity level of the fODF.  The fODF is a spherical signal in the \emph{HEALPix} sphere.   %

Another consideration of our method is that the determined fibre orientation is constrained to be antipodal symmetric using a real spherical harmonics transformation. Nevertheless, applying this characteristic to the estimated fODF produces the presence of negative coefficients. Since the fODF must always be positive in the unit sphere, it is necessary to constrain the non-negativity behaviour of the fODF. We assess that with the use of an $L_2$ regularisation over the non-positive values in the fODF defined as  $\mathcal{L}_N = \norm{\text{fODF}_{\text{fODF}_i < 0}}^2.$
Finally, the whole network can be trained end to end using the total loss $\mathcal{L}_{T}$ computed as $\mathcal{L}_{T} =
            \mathcal{L}_r +\mathcal{L}_{s} + \mathcal{L}_{N}.$

\vspace*{-.5\baselineskip}
\section{Results}%
\label{sec:results}
\vspace*{-.5\baselineskip}

\begin{figure*}[t]
	\includegraphics[width=\textwidth]{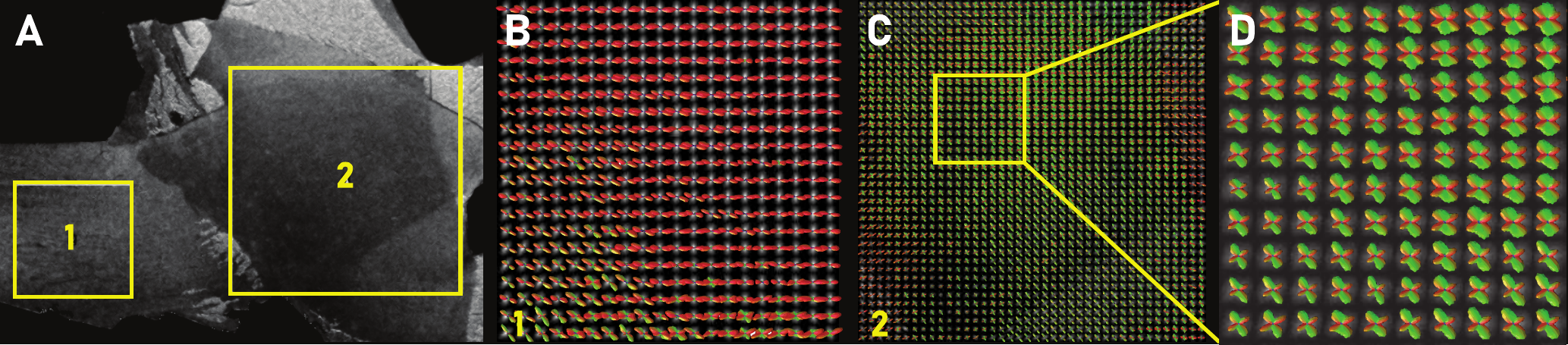}
	\caption{
	Three crossing sections of human optic tracts measured with scatterometry SLI. (A)Transmittance image. (B) fODF obtained  with GORDA superposed to its SLI measurement in a region of single fibres.(C-D) fODF obtained  with GORDA superposed to its SLI measurement in a region of crossing fibres. The fODF were visualized using MRVIEW~\cite{tournier2012mrtrix}.
	}
	\label{fig:odfs}
\end{figure*}

\textbf{Dataset.}
We tested the proposed approach using different postmortem tissue samples. The same tissue samples were used as in~\cite{menzel2021scattered, menzel2021scatterometry}: Three 30 \textmu m thin sections of human optic tracts that were manually placed on top of each other to obtain a model of three crossing nerve fibre bundles (sections no. 32/33), cf. Table F1 in~\cite{menzel2021scattered}. The optic tracts were extracted from the optic chiasm of a human brain (female body donor, 74 years old, without known neurological/psychiatric disorders) and cut along the fibre tracts of the visual pathway. Each scattering pattern was obtained by illuminating $2\times2$ LED in a LED panel composed of 256 LEDs, producing $81\times81$ pixels per pattern, providing a total of 19M SLI patterns for each section. In order to evaluate this study, a sub-sample of 10K patterns was randomly selected. 

\textbf{Training.}
The proposed unsupervised learning algorithm was optimized using AdamW~\cite{loshchilov2017decoupled}, with a learning rate of 0.01, batch size of $32$. Convergence was reached within 10-15 epochs, as expected given the design of \emph{Deepsphere}~\cite{perraudin2019deepsphere}. The training process was performed on a single node of JURECA-DC with one GPU (NVidia A100, 40GB). Our algorithm was implemented on top of Pytorch lightning~\cite{falcon2019pytorch}.
After training, the reconstruction model was disabled, and only the spherical U-Net was used to estimate fODF and further evaluate the method.

\textbf{Evaluation.} We evaluate the performance of the predicted fODF with the results obtained by SLIX in Table~\ref{tab:metrics}. We use the angular correlation coefficient (ACC), which measures the degree to which two functions over a sphere are correlated, and the Jensen-Shannon divergence (JSD) which measures the distance between two probability distributions~\cite{schilling2018histological}. We perform a separate analysis for regions containing single and crossing fibres in the optic tracts section.

\begin{table}[b]
\centering
\caption{Correlation and divergence scores for the comparison of GORDA with SLIX for regions with single and crossing fibres. The label $\uparrow$ implies the higher the better and $\downarrow$ implies the lower the better. }
\label{tab:metrics}
\begin{tabular}{@{}lcccc@{}}
\toprule
Region          & dataset                      & ACC 	$\uparrow$ & JSD $\downarrow$ \\ \midrule
Parallel fibres & \multirow{2}{*}{optic tracts} & 0.62 & 0.147\\
Crossing fibres &                              & 0.61 & 0.131    \\ 
\bottomrule
\end{tabular}
\end{table}

\tref{tab:metrics} shows a high correlation between GORDA and SLIX (when applied to the polar integral of the scattering patterns), which indicates that the results of our method can capture a similar fODF in the evaluated dataset. Fig \ref{fig:odfs} presents a qualitative impression of the performance of GORDA in the optic tracks sections. GORDA generates complete fODFs from the SLI scattering patterns for different fibre populations like single fibres as well as two and three crossing fibres.

\vspace*{-.5\baselineskip}
\section{Discussion \& Conclusion}%
\label{sub:conclusion}
\vspace*{-.5\baselineskip}

We introduced GORDA, a new strategy  to analyse SLI scattering patterns based on an unsupervised spherical convolutional network which provides a complete description of the fibre architecture. The calculated fODF is recovered without the need of any manual annotation or supervision and can be fully described on a spherical harmonics basis, which enables further comparison with other methodologies that calculate fODF in brain sections.
We presented a geometric model that enables the reconstruction of the scattering patterns by means of the fODF. Our approach opens a gateway to estimate the inclination of the reconstructed fODF from the SLI scattering patterns. 

Our experiments showed that the predicted fODF are capable of capturing different fibre mixtures, such as single fibres and crossing fibres. The correspondence of the estimated orientation was quantitatively compared with the analysis performed with SLIX. Remaining differences between the two methods result from 
(i) GORDA calculating a complete probabilistic function for the fODF, while SLIX estimates the peak orientation at each scattering pattern, and 
(ii) GORDA calculating the 3D orientation of the fibres, while SLIX detects the in-plane orientation of the fibres. 

In future work, we plan to perform an extensive evaluation for the inclination estimation of the scattering patterns compared with 3D-PLI scans; such comparison will enable to improve the parameters of our reconstruction model. Furthermore, we foresee the utility to perform knowledge distillation from SLIX  in the way of pseudo labels that could improve the overall performance of GORDA and improve the robustness of the estimation of the fODF.

\small{
\vspace*{\baselineskip}
\noindent{\textbf{Compliance with ethical standards}
The studies carried out require no ethical approvals.
The human postmortem brains for these studies were obtained by the body donor programs of the Anatomical Institute of University of Düsseldorf, Rostock and Aachen or from collaboration between Anatomical Institute and departments of pathology in accordance with legal requirements.
All body donors have signed a declaration of agreement.

\noindent{\textbf{Acknowledgments}
This project received funding from the European Union’s Horizon 2020 Research and Innovation Programme, grant agreement 945539 (HBP SGA3), and from the Helmholtz Association’s Initiative and Networking Fund through the Helmholtz International BigBrain Analytics and Learning Laboratory (HIBALL) under the Helmholtz International Lab grant agreement InterLabs-0015.
Computing time was granted through JARA on the supercomputer JURECA at Jülich Supercomputing Centre (JSC) as part of the project CJINM11.
The authors declare no competing interests.
}

\bibliography{refs}
\bibliographystyle{IEEEbib}

\end{document}